\documentclass[%
 reprint,
 amsmath,amssymb,
 aps,
 prx
]{revtex4-2}

\usepackage{amsfonts}
\usepackage{dsfont}
\usepackage{MnSymbol}
\usepackage{graphicx}
\usepackage{dcolumn}
\usepackage{xcolor}
\usepackage{physics}
\usepackage{bm}
\usepackage{float}
\usepackage{bbm}
\usepackage{ragged2e}
\usepackage[colorlinks=true,
		urlcolor=cyan,
            linkcolor=blue,
            anchorcolor=blue,
            citecolor=blue
            ]{hyperref}
\definecolor{darkred}{HTML}{bc4749}

\begin{document}

\preprint{APS/123-QED}

\title{Periodic conditional measurements on a photonic chip}

\author{Xuan Chen\textsuperscript{1,2}}
\author{Ruo-Yu Yin\textsuperscript{3,4}}%

\author{Xiao-Wen Shang\textsuperscript{1,2}}%
\author{Xi-Jun Yuan\textsuperscript{1,2}}%
\author{Ze-Feng Lan\textsuperscript{1,2}}%
\author{Hong-Jing Du\textsuperscript{1,2}}%
\author{Yi-Jun Chang\textsuperscript{1,2}}%
\author{Bao-Jing Liu\textsuperscript{1,2}}%
\author{Wen-Hao Zhou\textsuperscript{1,2}}%

\author{Xi Tan\textsuperscript{1,2}}%
\author{Eli Barkai\textsuperscript{3}}%
\author{Hao Tang\textsuperscript{1,2,$\dagger$}}%
\author{Xian-Min Jin\textsuperscript{1,2,5,6,$\dagger$}}%
 \affiliation{\textsuperscript{1}Center for Integrated Quantum Information Technologies (IQIT), School of Physics and Astronomy
and State Key Laboratory of Advanced Optical Communication Systems and Networks,
Shanghai Jiao Tong University, Shanghai 200240, China}
\affiliation{\textsuperscript{2}Hefei National Laboratory, Hefei 230088, China.}
\affiliation{\textsuperscript{3}Department of Physics, Institute of Nanotechnology and Advanced Materials, Bar-Ilan University, Ramat-Gan 52900, Israel}
\affiliation{\textsuperscript{4}Department of Physics \#1, Graduate School of Science, Kyoto University, Kyoto 606-8502, Japan}
\affiliation{\textsuperscript{5}TuringQ Co., Ltd., Shanghai 200240, China
}%

\affiliation{\textsuperscript{6}Chip Hub for Integrated Photonics Xplore (CHIPX), Shanghai Jiao Tong University, Wuxi 214000, China
}%

\date{\today}

\begin{abstract}
Quantum measurements not only extract information but also reshape quantum evolution through measurement backaction. Here, we investigate periodic conditional measurements, in which coherent evolution is repeatedly interrupted by local measurements. We show that the bright and dark modes of the conditional evolution operator provide a unified spectral description of both the post-selected final state along no-detection trajectories and the statistics of detection events. Using a three-dimensional photonic chip, we demonstrate on-demand state engineering through dark-mode preservation and bright-mode selection, enabling the protection of topological corner states and excited-state shelving. We further observe the predicted transition of the total detection probability on an effectively infinite one-dimensional lattice and identify its microscopic origin in Floquet near-dark modes. These results establish periodic conditional measurements as a general framework for engineering conditional quantum states and manipulating event-detection statistics in monitored quantum dynamics.
\end{abstract}
\maketitle

\section{Introduction}
Quantum measurements are traditionally used to extract information from quantum systems. Beyond this readout function, they can also serve as powerful tools for controlling quantum states and dynamics \cite{leggett2005quantum,pan2023weak,pan2020weak,rossini2020measurement,noel2022measurement,iadecola2023measurement, huo2026sequential}. Such measurement-based control can be realized through two conceptually distinct mechanisms. In adaptive protocols, measurement outcomes are used to determine subsequent operations, making the later evolution dependent on the acquired information \cite{gao2022quantum,spagnolo2022experimental,baldazzi2026coupled,raussendorf2003measurement,garcia2021learning}. Alternatively, a measurement can directly modify the system through measurement backaction, without requiring an explicit feedback operation, thereby enabling the investigation of nontrivial quantum states and dynamics \cite{hatridge2013quantum,groen2013partial,murch2008observation,muller2022measurement,koh2023measurement}. Among the various measurement protocols, conditional measurements provide a versatile way to harness measurement backaction for quantum control \cite{resch2002quantum,nakazato2003purification}. In such protocols, the state of a system is conditioned on a selected measurement outcome, often through post-selection, which projects the remaining degrees of freedom onto a desired quantum state or Hilbert subspace. This principle underlies a broad range of applications in quantum technologies, including photon addition and subtraction, quantum teleportation, entanglement purification, and probabilistic linear optical quantum gates \cite{magana2019multiphoton,dakna1998photon,bouwmeester1997experimental,pirandola2015advances,pan2001entanglement,liu2022optimal,knill2001scheme,carolan2015universal}.

The selected outcome, however, occurs only probabilistically. Most conditional measurement protocols focus on the state obtained from a single successful post-selection, while other outcomes are discarded. When the measurement is repeated periodically, consecutive no-click outcomes generate a well-defined conditional evolution through accumulated measurement backaction. A detection event instead terminates this evolution, and the measurement step at which it occurs varies randomly among experimental realizations, giving rise to a stochastic measurement record. Periodic conditional measurements thus connect two complementary aspects: the final state conditioned on the no-click branch and the statistics of when this branch terminates. Both are governed by the same survival operator $\mathcal{G}=(\mathbbm{1}-P)U(\tau)$.

Periodic conditional measurements make this shared dependence on $\mathcal{G}$ directly tunable by introducing the measurement period $\tau$. The coherent phases accumulated between successive measurements reshape the spectrum of $\mathcal{G}$, thereby determining both the state retained along the no-click branch and the temporal statistics of detection events. In this sense, periodic conditional measurements can be viewed as a non-unitary Floquet process, with $\mathcal{G}$ serving as the stroboscopic evolution operator over one measurement cycle \cite{shirley1965solution, sambe1973steady}. Varying $\tau$ can therefore generate non-unitary dynamics and spectral structures absent in single-shot protocols, opening opportunities for engineering conditional final states and tailoring detection statistics \cite{liu2022driving,singh2026transition,yin2019}. Experimental studies of periodic conditional measurements nevertheless remain scarce because their implementation requires repeated measurements and post-selection while preserving the coherent evolution of the surviving component.

Femtosecond-laser-written photonic waveguide arrays have emerged as a versatile platform for quantum information processing \cite{wang2023photonic,gao2022quantum,spagnolo2022experimental}, quantum computing \cite{li2022chip,maring2024versatile}, and quantum simulation \cite{tang2024simulating,lim2023oscillating}. The optical-quantum analogy maps paraxial light propagation onto Schrödinger evolution, enabling faithful simulations of quantum dynamics using classical light. Together with the flexible three-dimensional control of lattice geometries and coupling configurations, this platform has enabled large-scale two-dimensional quantum walks \cite{tang2018experimental}, non-Hermitian dynamics \cite{pan2018photonic,weidemann2022topological}, and Floquet engineering \cite{rechtsman2013photonic,yang2020photonic}. These capabilities make photonic waveguide arrays particularly well suited for investigating measurement-induced dynamics. Recent proposals for optical quantum control based on quantum Zeno dynamics \cite{liu2023engineering,liu2025measurement} further highlight their potential for exploring and harnessing measurement backaction.

Here, we realize periodic conditional measurements using a scalable, layer-structured three-dimensional photonic chip. Unlike the quantum Zeno limit, in which the measurement interval approaches zero, our architecture maps the propagation length between successive measurement layers directly onto a finite and precisely defined measurement period $\tau$. Within this platform, we investigate both the conditional final state and the measurement record through a unified spectral description based on the bright and dark modes of $\mathcal{G}$. For the final state, we demonstrate programmable state engineering by encoding target states into either the dark subspace or the longest-lived bright modes, enabling the protection of topological corner states and excited state shelving. For the measurement record, we investigate the statistics of detection events, experimentally observe the transition in the total detection probability on an effectively infinite one-dimensional lattice, and trace its microscopic origin to Floquet near-dark modes. These results establish periodic conditional measurements as a versatile framework for controlling post-selected states and probing monitored quantum dynamics.

\begin{figure*}[!pt]
\centering
\includegraphics[width=0.9\textwidth]{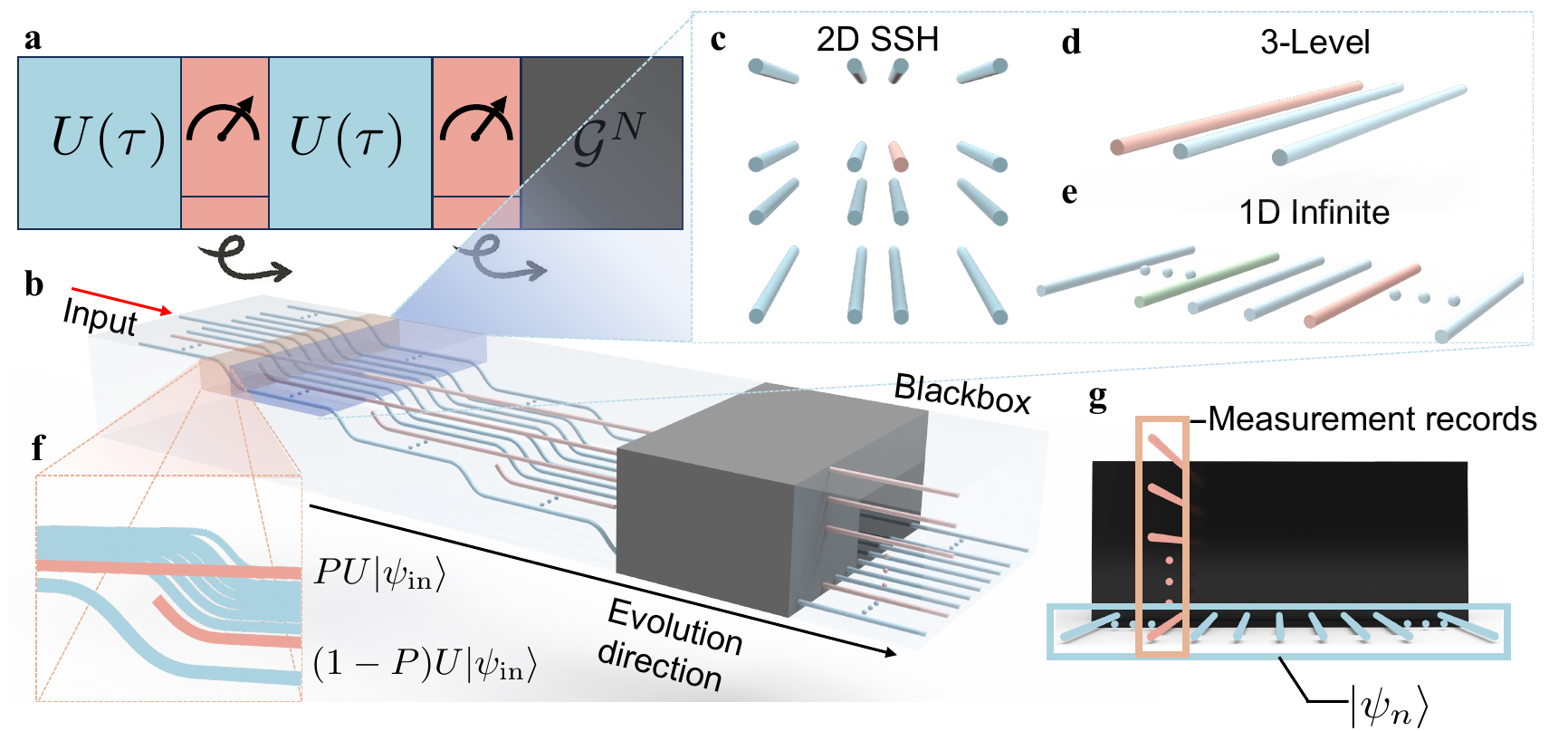}
\caption{
{\bf a,} Illustration of periodic conditional measurements. The coherent unitary evolution $U(\tau)$ is periodically interrupted by projective measurements, with the amplitude on the measured state removed from the system, thereby realizing conditional measurements.
The $n$ times repetition of this procedure is denoted by $\mathcal{G}^n$. 
{\bf b,} Snapshot of the photonic chip. 
The red waveguides represent the measured state and the blue waveguides represent the rest of the system. Each layer consists of an evolution phase (blue shaded area) followed by a conditional measurement (red shaded area) within a single period. The distance between two conditional measurements corresponds to the effective evolution duration $\tau$ within one period. The first four periods are exhibited while others are omitted in the black box.  {\bf c-e,} Waveguide structures adopted in our experiments. 
The waveguide corresponding to the measured state is colored red; 
the input waveguide of the effective infinite line is colored green, 
others are colored blue. {\bf f,} Detailed structure used to implement a conditional measurement. At the first measurement step, the component of the wavefunction on the measured state, $PU|\psi_{\rm in}\rangle$, is retained in the straight waveguide. At the corresponding position in the continuing system, an empty waveguide is introduced and joined with the other waveguides carrying the remaining amplitudes, thereby forming the post-selected no-detection state $(1-P)U|\psi_{\rm in}\rangle$.
{\bf g,} Output facet of the photonic chip. 
The measurement record (final state) is read from the waveguides framed with a red (blue) box.}
\label{fig:Fig1}
\end{figure*}

\section{Spectral modes and detection statistics of periodic conditional measurements}\label{model}
In this paper, the evolution of a quantum system under periodic conditional measurements is described by repeated application of the conditional evolution operator
\begin{equation}\label{eq1}
\mathcal{G} = (\mathbbm{1}-P)\,U(\tau),
\end{equation}
where $P=|\psi_{\rm m}\rangle\langle\psi_{\rm m}|$ is the projector onto the measured state 
$|\psi_{\rm m}\rangle$, 
$\mathbbm{1}$ is the identity operator, 
and $U(\tau)=e^{-iH\tau}$ is the unitary evolution between successive measurements separated by a period $\tau$ (setting $\hbar=1$),
governed by the system Hamiltonian $H$.
At every step,
the operator $\mathbbm{1}-P$ removes the population of the measured state, thereby postselecting the no-detection trajectories.
See Fig.~\ref{fig:Fig1}{\bf a} for a schematic illustration.

Starting from an initial state $|\psi_{\rm in}\rangle$, the conditional state after $n$ consecutive no-detection outcomes is
$|\psi_n\rangle=\mathcal{G}^n|\psi_{\rm in}\rangle/
\|\mathcal{G}^n|\psi_{\rm in}\rangle\|$.
Its long-time behavior determines the conditional final state investigated in Sec.~\ref{section:taloring}.

The same conditional evolution also determines the statistics of the measurement record. Across repeated realizations, a detection may occur at different measurement steps, thereby terminating the no-detection trajectory. The probability that the first detection occurs at the $n$th measurement is
$F_n=
\bigl|
\langle\psi_{\rm m}|U(\tau)\,
\mathcal G^{n-1}|\psi_{\rm in}\rangle
\bigr|^2$\cite{Kempe2005,Brun2006,Gruenbaum2013,Dhar2015,friedman2016quantum,yin2019,Wang2023pert,Giovanna2025causality,Sara2025}.
The resulting distribution $\{F_n\}$ characterizes the statistics of the measurement records and forms the basis of the analysis in Sec.~\ref{section:manipulation}.

Both the conditional final state and the statistics of the measurement record are governed by the spectrum and eigenmodes of the non-unitary operator $\mathcal{G}$, which depends jointly on the Hamiltonian $H$, the measured state $|\psi_{\rm m}\rangle$, and the measurement period $\tau$. We denote its right and left eigenmodes by $|\mu_j^{R}\rangle$ and $\langle\mu_j^{L}|$, respectively:
$
\mathcal{G}|\mu_j^{R}\rangle=\mu_j|\mu_j^{R}\rangle
$
and
$
\langle\mu_j^{L}|\mathcal{G}
=\mu_j\langle\mu_j^{L}|.
$
Here, $j$ labels the eigenmodes of $\mathcal G$, and the left and right eigenmodes are chosen to satisfy the biorthogonal normalization
$\langle\mu_j^{L}|\mu_k^{R}\rangle=\delta_{jk}$.

Expanding the initial state in this biorthogonal eigenbasis, the unnormalized surviving state after $n$ no-detection outcomes is~\cite{thiel2020dark,liu2022driving}
\begin{equation}\label{eq2}
|\tilde{\psi}_n\rangle
=
\sum_{j:\,0<|\mu_j|<1}
\zeta_j\,
\mu_j^n
|\mu_j^{R}\rangle
+
\sum_{j:\,|\mu_j|=1}
\zeta_j\,
\mu_j^n
|\mu_j^{R}\rangle ,
\end{equation}
where
$\zeta_j=\langle\mu_j^{L}|\psi_{\rm in}\rangle$
is the biorthogonal expansion coefficient. The corresponding normalized conditional state is
$
|\psi_n\rangle=
|\tilde{\psi}_n\rangle/
\sqrt{\langle\tilde{\psi}_n|\tilde{\psi}_n\rangle}.
$
The mode with $\mu_j=0$, which vanishes after the first application of $\mathcal G$, has been omitted from Eq.~\eqref{eq2}.

The two terms in Eq.~\eqref{eq2} reveal two distinct classes of conditional eigenmodes. We denote the bright-mode subspace by
$
\mathcal B=\{|\mu_j^R\rangle:0<|\mu_j|<1\}
$
and the dark-mode subspace by
$
\mathcal D=\{|\mu_j^R\rangle:|\mu_j|=1\}.
$
A bright mode develops a nonzero amplitude on the measured state during the unitary evolution. Each subsequent measurement removes part of this amplitude, so its surviving weight decreases as $|\mu_j|^{2n}$ and the removed probability contributes to the measurement record. A dark mode, in contrast, has exactly zero amplitude on the measured state immediately before every measurement. It is therefore invisible to the detector, retains its norm, and acquires only a phase after each measurement cycle.

This spectral separation determines the long-time conditional state. If the initial state has a nonzero component in $\mathcal D$, repeated measurements progressively remove the components in $\mathcal B$, making the dark component increasingly prominent in the normalized surviving state. The system consequently approaches the populated dark subspace. If the initial state has no component in $\mathcal D$, repeated measurements preferentially suppress the faster-decaying bright modes, and the conditional state approaches the longest-lived bright mode with the largest $|\mu_j|$.

The physical origin of a dark mode is the absence of amplitude on the measured state at each measurement step. The simplest realization is an energy eigenstate $|E\rangle$ satisfying
$\langle\psi_{\rm m}|E\rangle=0$.
Its spatial profile already avoids the measured state, and the measurement therefore has no effect on its evolution. Dark modes can also arise through destructive interference. When several energy eigenstates acquire the same phase over one measurement period, an appropriate superposition can make their amplitudes cancel exactly on the measured state. Because their relative phases are preserved from one measurement step to the next, this cancellation persists throughout the conditional evolution. The required phase degeneracy may be intrinsic, as for degenerate energies, or induced by the measurement period through
$
e^{-iE_1\tau}=e^{-iE_2\tau},
$
equivalently
$
\tau|E_1-E_2|=2m\pi.
$
Tuning $\tau$ can therefore create or reconstruct dark and near-dark modes without modifying the underlying Hamiltonian.

\section{Experimental realization}
To experimentally implement periodic conditional measurements, we develop a three-dimensional layer-structured photonic chip fabricated in borosilicate glass by femtosecond laser direct writing (Fig.~\ref{fig:Fig1}{\bf b}). Light propagation in the waveguide array obeys the Schrödinger-type equation $i\partial_z\psi=H\psi$, with the propagation distance $z$ mapping onto an equivalent evolution time. This optical-quantum correspondence enables the monitored quantum dynamics to be faithfully simulated using coherent classical light. The waveguide array is described by the tight-binding Hamiltonian
\begin{equation}
    H=\sum_{<i,j>}(J_{ij}|i\rangle\langle j| +H.c.)+\sum_j\Delta\beta_j |j\rangle\langle j|,
\end{equation}
Here, \(|j\rangle\) denotes the state localized at waveguide \(j\), \(J_{ij}\) is the nearest-neighbor coupling strength between waveguides \(i\) and \(j\), and \(\Delta\beta_j\) is the propagation-constant detuning of waveguide \(j\) relative to a common reference, which implements the corresponding on-site energy. The detuning is controlled through the femtosecond-laser writing power, whereas the coupling strength is determined primarily by the waveguide separation. Longer-range couplings are neglected because they decay exponentially with separation. By engineering the waveguide arrangement, coupling strengths, and on-site detunings, we implement the different Hamiltonians shown in Fig.~\ref{fig:Fig1}{\bf c--e} to investigate distinct conditional dynamics.

The waveguide array is divided into successive layers, each consisting of a coherent evolution stage followed by a conditional measurement. The propagation length of each evolution stage, shown by the blue shaded region in Fig.~\ref{fig:Fig1}{\bf b}, directly determines the measurement period. The conditional measurement operator $\mathbbm{1}-P$ is implemented by routing all unmeasured waveguides into the next layer while leaving the waveguide corresponding to the measured state straight (red shaded region). A vertical separation between adjacent layers strongly suppresses their mutual coupling as illustrated in Fig.~\ref{fig:Fig1}{\bf f}. Consequently, the amplitude occupying the measured state remains in the straight waveguide and propagates independently to the output, whereas the remaining field enters the next layer with an empty measured site and continues to evolve coherently under the same Hamiltonian. This removes the measured-state component from the evolving wavefunction and realizes the operation $\mathbbm{1}-P$.

\begin{figure*}[t]
\includegraphics[width=0.9\textwidth]{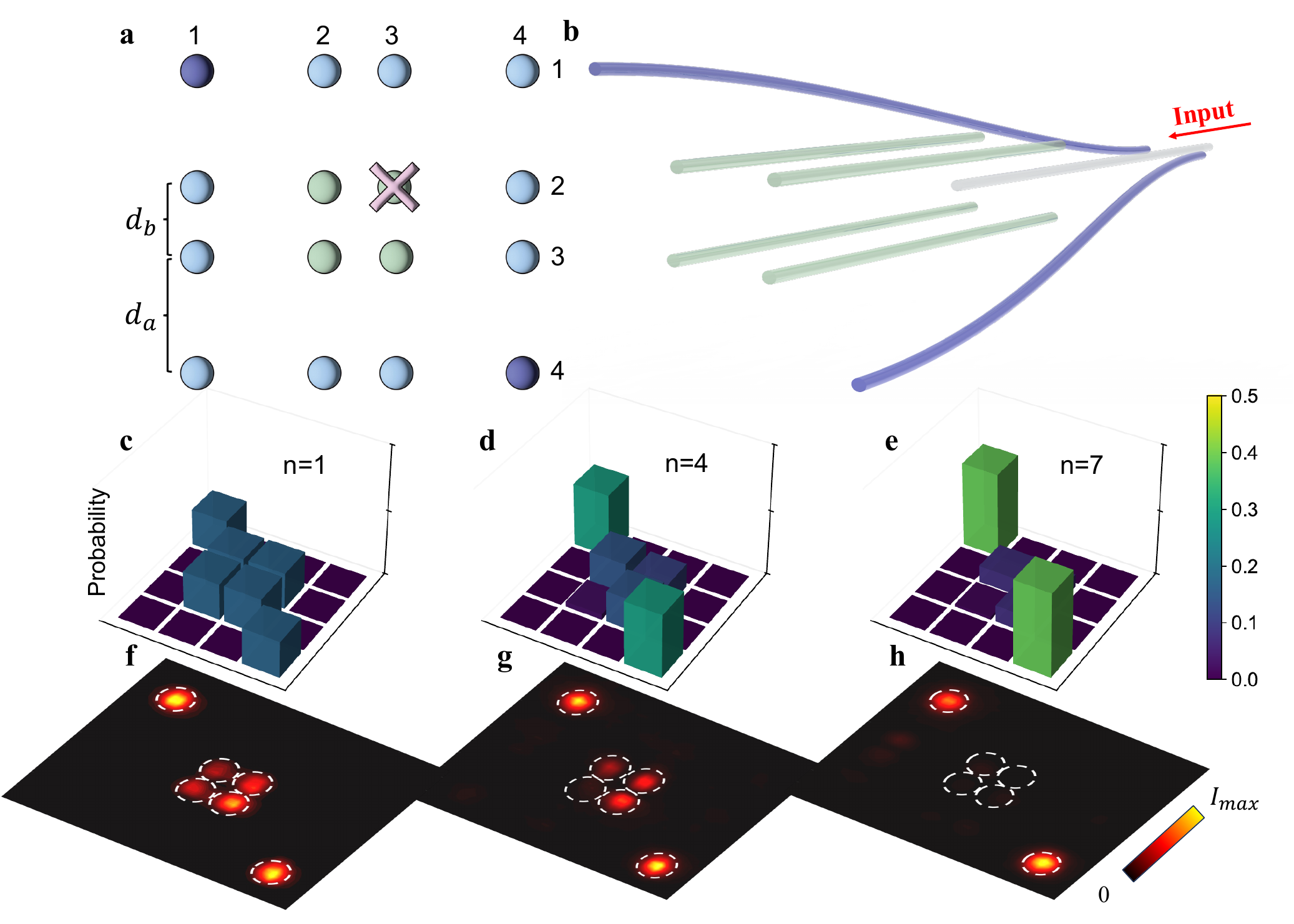}
\caption{ {\bf a,} Illustration of the 2D SSH lattice used in our experiment. The system contains 16 sequentially indexed sites. The intracell and intercell coupling strengths are determined by the waveguide separations $d_a$ and $d_b$, respectively. The injection sites associated with the selected corner and bulk states are colored blue and green, respectively [see also Eq.~(\ref{eq6})]. Periodic conditional measurements are performed on the state $|2,3\rangle$, which has a nonzero overlap with the input bulk state. {\bf b,} Waveguide structure for input-state preparation. Light injected into the central waveguide sequentially passes through a $1\times3$ and a $1\times4$ coupler to generate the selected corner and bulk components, respectively. The two components are then simultaneously injected into the SSH lattice to prepare their superposition. For the results shown here, the representative amplitude ratio ($\sqrt{2}\alpha=\beta$) is used. {\bf c--e,} Simulated probability distributions immediately before conditional measurement steps $n=1,4,$ and $7$. {\bf f--h,} Corresponding CCD-captured intensity distributions at $n=1,4,$ and $7$.}
\label{fig:Fig2}
\end{figure*}

Repeating this layer structure implements alternating coherent evolutions and conditional measurements. The writing depth and precision of the photonic chip provide scalability in the Hilbert-space dimension and number of measurement steps, while the evolution length allows flexible control of the measurement period. The first few periods are illustrated in Fig.~\ref{fig:Fig1}\textbf{b}; the remaining periods are omitted in the blackbox for clarity.

At the output facet (Fig.~\ref{fig:Fig1}\textbf{g}), the intensity in each waveguide is recorded using a charge-coupled device (CCD) camera. The lattice-waveguide outputs in the blue box give the probability distribution of the conditional final state, $|\langle j|\psi_n\rangle|^2$. The straight measurement waveguides in the red box record the first-detection statistics: the intensity in the waveguide from the $n$th layer gives the probability that the system survives the first $n-1$ measurements and is first detected in state $|\psi_{\rm m}\rangle$ at the $n$th measurement, corresponding directly to $F_n$. Collecting the intensities from all measurement waveguides therefore yields the complete measurement record.

The platform thus provides simultaneous access to the two complementary aspects of periodic conditional measurements: the post-selected conditional final state and the first-detection statistics of the measurement record. The following sections show how the conditional spectrum governs both programmable state engineering and repeated detection events.

\section{quantum state engineering with dark modes and bright modes}
\label{section:taloring}

In this section, we demonstrate quantum state engineering through dark-mode preservation and bright-mode selection under periodic conditional measurements.

\textit{Dark modes---}We first consider a two-dimensional SSH lattice in its topologically nontrivial phase~\cite{dutt2020higher,lustig2022photonic}, described by
\begin{equation}
\begin{aligned}
H
=&\sum_{m,n}J_x(m)\,|m+1,n\rangle\langle m,n|\\
&+\sum_{m,n}J_y(n)\,|m,n+1\rangle\langle m,n|
+\mathrm{H.c.},
\end{aligned}
\end{equation}
where $|m,n\rangle$ denotes the state localized at lattice site $(m,n)$. The Hamiltonian is realized in a $4\times4$ waveguide array with alternating separations $d_a=24~\mu\mathrm{m}$ and $d_b=11~\mu\mathrm{m}$, which determine the intracell and intercell coupling strengths, respectively (Fig.~\ref{fig:Fig2}{\bf a}). Because the intercell coupling is stronger than the intracell coupling, the lattice lies in the topologically nontrivial regime and supports modes localized near its four corners~\cite{xie2018second,wang2021quantum}.

We apply conditional measurements to the bulk site $(2,3)$, corresponding to the measured state $|\psi_{\rm m}\rangle=|2,3\rangle$. Owing to their spatial localization, the corner modes have zero amplitude on this site and are orthogonal to $|\psi_{\rm m}\rangle$, thereby forming a dark subspace of $\mathcal G$. This orthogonality condition holds independently of the measurement period.

To demonstrate the resulting state selection, we prepare an input containing a corner-localized component and a component excited on the four central sites,
\begin{equation}\label{eq6}
\begin{aligned}
|\psi_{\rm in}\rangle
&=\alpha|\psi_{\rm c}\rangle+\beta|\psi_{\rm b}\rangle,\\
|\psi_{\rm c}\rangle
&=\frac{1}{\sqrt{2}}
\left(
|1,1\rangle+|4,4\rangle
\right),\\
|\psi_{\rm b}\rangle
&=\frac{1}{2}
\left(
|2,2\rangle+|2,3\rangle
+|3,2\rangle+|3,3\rangle
\right),\\
R
&=\mathbbm{1}-|\psi_{\rm m}\rangle\langle\psi_{\rm m}|\\
&=\mathbbm{1}-|2,3\rangle\langle2,3|.
\end{aligned}
\end{equation}
Here, $|\psi_{\rm c}\rangle$ is a superposition of two diagonally opposite corner-localized modes, with nonzero coefficients $\zeta_j$ only for the dark modes in $\mathcal{D}$. In contrast, $|\psi_{\rm b}\rangle$ is a uniform excitation of the four central sites and has substantial weights on the bright modes in $\mathcal{B}$. Repeated measurements suppress these bright components while retaining $|\psi_{\rm c}\rangle$. We choose $\tau=5000~\mu\mathrm{m}$ for the experimental demonstration.

\begin{figure*}[t]
\includegraphics[width=1\textwidth]{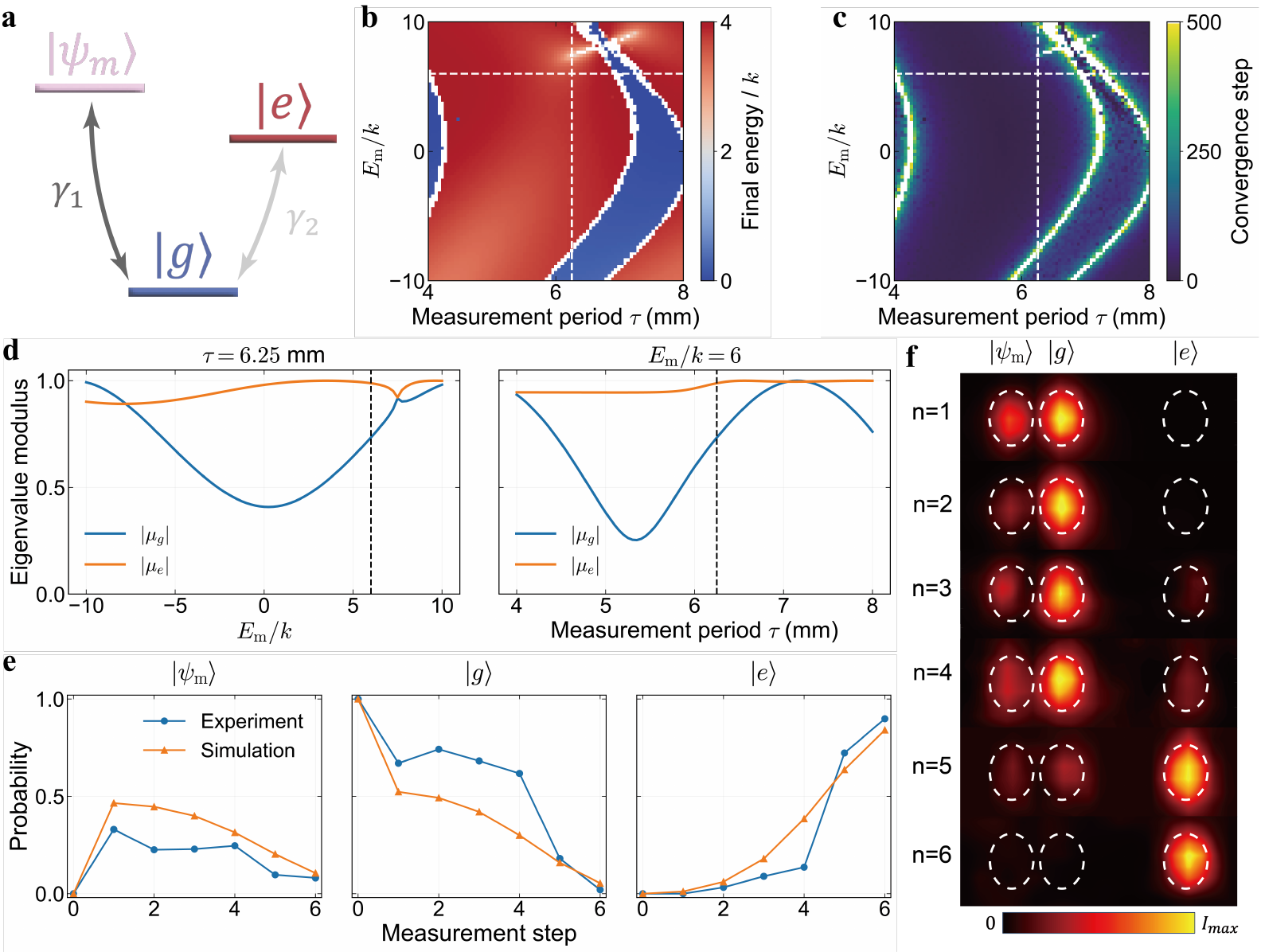}
\caption{
{\bf a,} Illustration of the three-level system. The ground state $|g\rangle$ is coupled to the measured state $|\psi_{\rm m}\rangle$ with strength $\gamma_1$ and to the excited state $|e\rangle$ with strength $\gamma_2$.
{\bf b,c,} Final energy and convergence step as functions of the measured-state energy and measurement period. The final energy converges to either the ground- or excited-state energy. The boundary corresponds to nearly degenerate moduli of the two dominant eigenvalues, where the convergence step diverges.
{\bf d,} Moduli of the $|g\rangle$- and $|e\rangle$-dominated eigenvalues, $|\mu_g|$ and $|\mu_e|$ when:
(i) fixing the measurement period at $\tau=6.25$ $\rm{mm}$ and scanning the measured-state energy, and
(ii) fixing the measured-state energy at $E_{\rm m}/k=6$ and scanning the measurement period.
{\bf e,} Simulated and experimental occupation probabilities immediately before each conditional measurement.
{\bf f,} CCD-captured intensity distributions immediately before each conditional measurement.}
\label{fig:Fig3}
\end{figure*}

The simulated probability distributions immediately before measurement steps \(n=1,4,\) and \(7\) are shown in Fig.~\ref{fig:Fig2}{\bf c--e}. The bulk-state component, which projects onto bright modes of \(\mathcal G\), is progressively suppressed, whereas the corner-state component encoded predominantly in the dark subspace is retained. The corresponding CCD measurements in Fig.~\ref{fig:Fig2}{\bf f--h} agree well with the simulations and show that the bulk-state contribution is effectively eliminated after seven conditional measurements. This agreement validates our photonic implementation of the conditional evolution and confirms that the designed measurement sequence accurately realizes the expected dark- and bright-mode dynamics. 

\textit{Bright modes---}Beyond the preservation of dark modes, bright modes can also serve as a resource for state engineering. Because different bright modes decay at different rates under repeated conditional measurements, the conditional evolution can selectively retain the longest-lived one and thereby steer the normalized surviving state toward a desired target. We demonstrate this complementary strategy using a V-shaped three-level system.
\begin{equation}
    \begin{gathered}
    H=E_{\rm m}|\psi_{\rm m}\rangle\langle \psi_{\rm m}|+E_g|g\rangle\langle g|+E_e|e\rangle\langle e|+\\\gamma_1(|\psi_{\rm m}\rangle\langle g|+|g\rangle\langle \psi_{\rm m}|)+\gamma_2(|g\rangle\langle e|+|e\rangle\langle g|),
    \end{gathered}    
\end{equation}
where $|\psi_{\rm m}\rangle$, $|g\rangle$, and $|e\rangle$ denote the measured, ground, and excited states, respectively; $E_{\rm m}$, $E_g$, and $E_e$ are their on-site energies; $\gamma_1$ and $\gamma_2$ are the corresponding coupling strengths (Fig.~\ref{fig:Fig3}{\bf a}). The Hamiltonian is mapped onto three waveguides whose propagation constants and separations control the on-site energies and coupling strengths, respectively (Fig.~\ref{fig:Fig1}{\bf d}).

With conditional measurements applied to $|\psi_{\rm m}\rangle$, the corresponding operator ${\cal G}$ has three eigenmodes, one of which has a zero eigenvalue, while the other two belong to the bright subspace $\mathcal{B}$. In this case, the dark subspace $\mathcal{D}$ is empty. When the moduli of the two bright eigenvalues are nondegenerate, the normalized long-time state approaches the eigenmode with the largest modulus, provided that it has a nonzero overlap $\zeta_j$ with the initial state.

We set the coupling strengths $\gamma_{1}$ and $\gamma_{2}$ by choosing waveguide separations of $8~\mu\mathrm{m}$ and $20~\mu\mathrm{m}$, respectively, such that $\gamma_{1}\gg\gamma_{2}$, and set the on-site energies to $E_g=0$ and $E_e=4k$, where $k$ denotes the propagation constant change induced by a $1~\mathrm{mW}$ increase in the power of laser direct writing. Under this configuration, the two bright eigenmodes of $\mathcal G$ are close to $|g\rangle$ and $|e\rangle$, respectively, and are therefore referred to as the $|g\rangle$- and $|e\rangle$-dominated bright modes, with corresponding eigenvalues denoted by $\mu_g$ and $\mu_e$. Since \(\gamma_{1}\gg\gamma_{2}\), \(|g\rangle\) couples more strongly to the measured state and might therefore be expected to decay more rapidly under repeated conditional measurements, leaving \(|e\rangle\) as the surviving state. However, the relative lifetimes of the bright modes are determined by the full conditional spectrum and depend additionally on the measured-state energy and the measurement period.

Fig.~\ref{fig:Fig3}\textbf{b} shows the final energy of the system as a function of $E_{\rm m}/k$ and $\tau$. The parameter space separates into two regions in which the final energy approaches either $4k$, corresponding to $|e\rangle$, or $0$, corresponding to $|g\rangle$. Near the boundary, the moduli of the two dominant eigenvalues become nearly degenerate, causing the number of measurements required for convergence to increase sharply. The corresponding convergence steps are shown in Fig.~\ref{fig:Fig3}\textbf{c}; region with measurement steps over 500 are treated as divergent and left blank. We further scan $E_{\rm m}$ at fixed $\tau=6.25~\mathrm{mm}$ and scan $\tau$ at fixed $E_{\rm m}=6k$, corresponding to the dashed lines in Fig.~\ref{fig:Fig3}\textbf{b-d}. Along both scans, $|\mu_g|$ and $|\mu_e|$ first approach each other near the nonconvergent boundary and then exchange their ordering (Fig.~\ref{fig:Fig3}\textbf{d}). This exchange switches the longest-lived bright mode, causing the conditional evolution to converge to different final states on the two sides of the boundary.

For the experiment, we choose the intersection of these two scans, and initialize the system in $|\psi_{\rm in}\rangle=|g\rangle$. The simulated and experimental occupation probabilities immediately before each conditional measurement are shown in Fig.~\ref{fig:Fig3}\textbf{e}, with experimental results represented by blue circles and simulations by orange triangles. They correspond to $|\langle\phi|U(\tau){\cal G}^{n-1}|\psi_{\rm in}\rangle|^2$, where $|\phi\rangle=|\psi_{\rm m}\rangle,|g\rangle,|e\rangle$.

As the number of measurements increases, the populations in $|\psi_{\rm m}\rangle$ and $|g\rangle$ decrease, while that in $|e\rangle$ progressively increases in the normalized conditional state. After five measurements, the surviving intensity is predominantly localized in $|e\rangle$, as confirmed by the CCD image in Fig.~\ref{fig:Fig3}\textbf{f}. This evolution reflects the differential decay of the bright modes: components associated with smaller $|\mu|$ are removed more rapidly, leaving the longest-lived bright mode, which is engineered here to be dominated by $|e\rangle$.

Together with the SSH-lattice results, these observations demonstrate two complementary routes to state engineering under periodic conditional measurements: preserving a target component in the dark subspace or selecting it as the longest-lived bright mode. More broadly, the target conditional state is determined jointly by the Hamiltonian, the measured state, and the measurement period. The latter two provide additional control parameters that can reshape the conditional spectrum without modifying the intrinsic Hamiltonian, establishing periodic conditional measurements as an alternative approach to quantum-state control.

\section{Statistics of Measurement records under periodic conditional measurements}
\label{section:manipulation}

In monitored quantum systems, the sequence of detection and no-detection outcomes constitutes a stochastic measurement record, whose statistics reveal when and whether a designated event occurs. Within the quantum-trajectory formalism, such records arise from the measurement conditioned evolution of the system~\cite{carmichael1993open,dalibard1992wave,plenio1998quantum}. In continuously monitored systems, the record and the associated backaction are governed primarily by the measurement strength or monitoring rate, enabling, for example, the real-time tracking and control of quantum jumps~\cite{minev2019catch}. By contrast, periodic conditional measurements interrogate the system projectively at discrete intervals $\tau$. Tuning $\tau$ can create or reconstruct Floquet dark and near-dark modes in the spectrum of $\mathcal G$, thereby strongly modifying the detection statistics.

\begin{figure*}[!tp]
\includegraphics[width=1.0\textwidth]{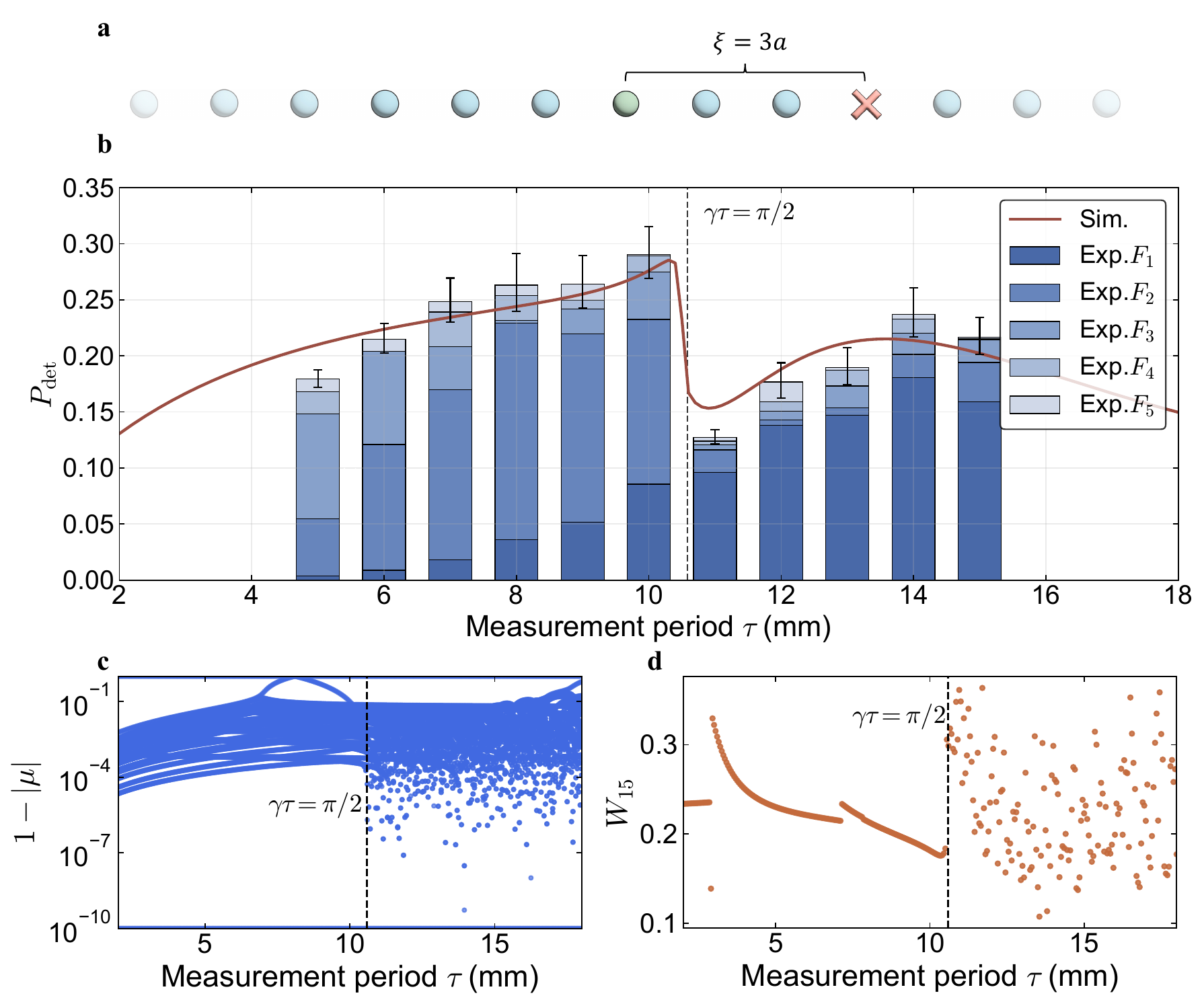}
\caption{{\bf a,} Schematic of the monitored quantum walk on a one-dimensional chain. The walker is initialized at the central site colored green, while periodic conditional measurements are performed at a target site located three lattice constants away $\xi=3a$ with measurement period $\tau$.
{\bf b,} Experimental total detection probability $P_{\rm{det}}$ as a function of the measurement period. The stacked bars represent the contributions of the first five terms of the first-detection distribution first-detection probabilities from $F_1$ to $F_5$, and the red curve is the theoretical prediction. A sharp transition is observed around the critical measurement period $\gamma\tau=\pi/2$ (vertical dashed line), where the total detection probability reaches its maximum. 
{\bf c,} 
Spectrum of the conditional evolution operator $\mathcal{G}$, plotted as $1-|\mu|$ versus the measurement period. Around $\tau_c$, a manifold of Floquet near-dark modes with eigenvalues approaching the unit circle emerges. For larger measurement periods, the conditional spectrum is continuously reconstructed as additional Floquet quasi-degeneracies develop.
{\bf d,} 
Cumulative overlap $W_{15}$ between the eigenstates with 15 largest $|\mu|$ and the initial state. The abrupt increase of $W_{15}$ around $\tau_c$ indicates that the initial state acquires substantially larger overlap with the emerging Floquet near-dark modes. Beyond the transition, the continuous reconstruction of the conditional spectrum leads to the oscillatory behavior of $W_{15}$ as well as $P_{\rm{det}}$.
}
\label{fig:Fig4}
\end{figure*}

Here, we characterize the statistics of the measurement record using the first-detection probability $F_n$, which specifies the probability that the measured state is first detected at the $n$th measurement as defined in Sec.\ref{model}, and the total detection probability $P_{\rm det}$, which determines whether it is eventually detected~\cite{Kempe2005,Brun2006,Gruenbaum2013,Dhar2015,friedman2016quantum,thiel2020dark}.

\begin{equation}\label{eq:Pdet}
P_{\rm det}\equiv\sum_{n=1}^{\infty}F_n=
1-\sum_{|\mu_j\rangle\in \mathcal{D}}|\zeta_j|^2 .
\end{equation}
The last equality follows from the spectral decomposition of $\mathcal G$. Components associated with the bright modes $|\mu_j\rangle\in B$ decay under repeated measurements and eventually contribute to the detection probability, whereas the initial weight in the dark modes remains permanently undetected. A nonzero dark-mode weight therefore leads to incomplete detection, $P_{\rm det}<1$~\cite{Brun2006,thiel2020dark}. The detection statistics are thus governed by the bright- and dark-mode structure of the conditional dynamics and can be controlled by reshaping this structure through the measurement period.

Typically, obtaining the statistics of the measurement record requires a large ensemble of realizations and repeated measurements
\cite{tornow2023measurement,wang2024first,yin2024b}.
However, in our photonic platform, the ensemble-averaged measurement record is directly encoded in the output intensities. Denoting the intensity in the $n$th measurement-record waveguide by $I_n$ and that in the final-state waveguides by $I_{\rm final}$, the first-detection probability is obtained as
$F_n=I_n/I_{\rm total}$, where $I_{\rm total}=\sum_n I_n+I_{\rm final}$ is the total output intensity. The total detection probability is then given by $P_{\rm det}=\sum_n F_n$. Thus, both $F_n$ and $P_{\rm det}$ can be obtained from a single spatially resolved output image for each experimental setting, without sampling individual detection events over repeated state preparations.

To investigate how the measurement period shapes the statistics of the measurement record, we consider first detection in a continuous-time quantum walk on a one-dimensional lattice~\cite{ben2004one}. The Hamiltonian is
\begin{equation}\label{Ham_Line}
H=-\gamma\sum_{x\in a\mathbb{Z}}
\left(
|x\rangle\langle x+a|
+
|x+a\rangle\langle x|
\right),
\end{equation}
where $a$ is the lattice constant and $\gamma$ denotes the nearest-neighbor coupling strength. The walker is initialized at the central lattice site, while the designated measurement event is defined as the projection onto a site located three lattice constants $\xi=3a$ away from the initial position, as illustrated in Fig.~\ref{fig:Fig4}\textbf{a}. Repeated conditional measurements are performed at this target site with a measurement period $\tau$. Experimentally, the quantum walk is implemented using an array of 61 waveguides with a uniform spacing of $14.1~\mu\mathrm{m}$, corresponding to a nearest-neighbor coupling strength of $\gamma\simeq0.148~\mathrm{mm}^{-1}$. The central waveguide is excited to prepare the initial state.

This setting captures the competition between coherent transport and periodic interrogation.
When measurements are too frequent, transport toward the measured site is inhibited by the quantum Zeno effect \cite{itano1990quantum}; when they are too sparse, the wavepacket may pass the measured site between successive interrogations. The competition between these two limits suggests the existence of an optimal measurement period. For an infinite one-dimensional chain, Ref.~\cite{Thiel2018PRL} predicted a critical period $\gamma\tau_c=\pi/2$, at which $P_{\rm det}$ reaches its maximum value and its dependence on the measurement period changes from a monotonic increase to an oscillatory decay. The corresponding theoretical prediction is shown by the red curve in Fig.~\ref{fig:Fig4}\textbf{b} and serves as the benchmark for our experiment.

Our experiment is implemented on a finite lattice rather than a strictly infinite chain. Nevertheless, for the measurement periods and number of attempts considered here, the wavepacket remains well separated from the boundaries, allowing the system to faithfully reproduce the infinite-chain dynamics. This effective realization permits direct comparison with the theoretical prediction while offering an additional advantage: the finite conditional evolution operator $\mathcal{G}$ can be diagonalized, providing a tractable spectral approach to understanding the transition predicted for the infinite system. In particular, eigenmodes with $|\mu|$ close to unity lose little amplitude under each measurement and thus behave as near-dark modes. By examining how their eigenvalues and overlaps with the initial state vary with the measurement period, we can identify the spectral origin of the observed behavior of $P_{\rm det}$.

In our experiment, the first five terms of the first-detection distribution, $F_1$--$F_5$, are sufficient to estimate $P_{\rm det}$, since the remaining contribution is negligible owing to the rapid decay of $F_n$ \cite{Thiel2018PRL}. We fabricate samples with measurement periods ranging from 5 to 15 mm in 1 mm increments and compare the experimentally measured $P_{\rm det}$ with the theoretical prediction. 
As shown in Fig.~\ref{fig:Fig4}\textbf{b}, a transition of the total detection probability is experimentally observed around the critical measurement period $\tau_c=\pi/2\gamma\simeq10.6$~mm, separating a monotonic regime from a nonmonotonic regime. 

To reveal the microscopic origin of the transition, we analyze the spectrum of the finite conditional evolution operator $\mathcal G$. For short measurement periods, increasing $\tau$ allows the wavepacket to propagate farther between successive measurements, resulting in a monotonic increase of $P_{\rm det}$. For the infinite tight-binding chain described by Eq.~\eqref{Ham_Line}, the dispersion relation is $E(k)=-2\gamma\cos(ak)$, with band edges $E_{\rm top}=2\gamma$ and $E_{\rm bottom}=-2\gamma$ and a bandwidth of $4\gamma$~\cite{ben2004one}. At the critical period $\gamma\tau_c=\pi/2$, the phases accumulated at the two band edges coincide because
$
(E_{\rm top}-E_{\rm bottom})\tau_c
=4\gamma\tau_c=2\pi.
$
This condition marks the first Floquet quasi-degeneracy spanning the full energy band. In the finite lattice, the discrete spectrum slightly lifts the exact degeneracy, but superpositions of the nearly phase-degenerate components can develop strongly suppressed amplitudes on the measured site. A group of long-lived Floquet near-dark modes consequently emerges, with eigenvalues $\mu_j$ clustered near the unit circle, $|\mu_j|\simeq1$, as shown in Fig.~\ref{fig:Fig4}\textbf{c}. As $\tau$ increases beyond $\tau_c$, additional Floquet quasi-degeneracies emerge within the band and continuously reorganize the near-dark-mode structure, leading to the nonmonotonic dependence of $P_{\rm det}$ on the measurement period.

In addition to their eigenvalue moduli, the influence of the emerging near-dark modes on $P_{\rm det}$ depends on their overlap with the initial state, as indicated by Eq.~\eqref{eq:Pdet}. Ordering the modes as $|\mu_1|\geq|\mu_2|\geq\cdots$, we define the cumulative weight of the $k$ longest-lived modes as
$
W_k=\sum_{j=1}^{k}|\zeta_j|^2
=\sum_{j=1}^{k}
\left|\langle\mu_j^L|\psi_{\rm in}\rangle\right|^2.
$
As shown in Fig.~\ref{fig:Fig4}\textbf{d}, $W_{15}$ increases abruptly near $\tau_c$, indicating that the initial state acquires substantial weight in the emerging near-dark modes. For $\tau>\tau_c$, the repeated reconstruction of these modes produces corresponding variations in $W_{15}$ and $P_{\rm det}$.

Although the transition was originally predicted for an infinite lattice, our finite-dimensional spectral framework captures its microscopic origin and can be readily generalized to a broad class of lattice models with different band structures. More importantly, it identifies the measurement period as a key control parameter of monitored quantum dynamics, providing a practical guideline for tailoring measurements either to maximize the probability of detecting a target event or to minimize measurement backaction and preserve the target state.

\section{Discussion}
In this work, we establish a unified framework for periodic conditional measurements and demonstrate it experimentally on a three-dimensional photonic chip. The deterministic evolution along the post-selected no-click branch and its stochastic termination by a detection event are complementary aspects of the same measurement process. Together with our layered photonic-chip design, this perspective connects conditional state engineering and quantum first detection through the spectrum of the same non-unitary conditional evolution operator $\mathcal G$.

In addition to the Hamiltonian, the spectrum and eigenmodes of $\mathcal G$ depend on the measured state and measurement period, providing additional parameters for state engineering without modifying the intrinsic system. We demonstrate two corresponding routes for selecting a desired state: preserving it as a dark mode or engineering it as the longest-lived bright mode. The role of the measurement period is particularly distinct from that in the quantum Zeno limit, where increasingly frequent measurements suppress transitions and confine the dynamics to a Zeno subspace\cite{liu2023engineering,liu2025measurement}. Here, a finite period allows coherent evolution to accumulate between successive measurements, so that varying $\tau$ reshapes the conditional spectrum and changes the selected final state.

The same conditional spectrum also governs the measurement record. On an effectively infinite one-dimensional lattice, we observe the predicted transition in the total detection probability and trace it to Floquet near-dark modes emerging when the measurement period brings the two band edges into quasienergy resonance~\cite{Thiel2018PRL}. This establishes a connection between detection statistics and characteristic features of the underlying energy band. Extending this approach to higher-dimensional lattices may provide a means of analyzing more complex band structures through their measurement-period-dependent detection response, while guiding the design of measurement strategies that enhance or suppress event detection.

The photonic implementation also provides a practical basis for extending this framework beyond the present fixed measurement configuration. Future integration with electro-optic control could enable reconfigurable and adaptive evolution between successive measurements, allowing the measured state, measurement period, and intermediate unitary dynamics to be adjusted during the process. Beyond enlarging the accessible conditional spectra, such programmability would allow measurement sequences to be designed for specific tasks, offering enhanced control over the preparation of target states and the tailoring of detection profiles. These capabilities would establish integrated photonics as a programmable platform for exploring and controlling measurement-driven quantum dynamics.

\section*{Methods}
\subsection*{Femtosecond laser direct written waveguides}
The fabrication of waveguides is realized by focusing a femtosecond laser beam to change the refractive index in a 100 $\rm{mm}$ $\times$ 20 $\rm{mm}$ borosilicate substrate. The focus point can be moved by accurate manipulation of the 3D motion stage to construct a continuous waveguide structure where single-mode light of 810 $\rm{nm}$ can be confined in the center and propagate. The femtosecond laser has a repetition rate of 1MHz, central wavelength 513 $\rm{nm}$, pulse duration 290 $\rm{fs}$, and pulse energy 190 $\rm{nJ}$, and is modulated by a Spatial Light Modulator (SLM) and then focused by a 50× objective lens. The moving speed of the focus point is 5 $\rm{mm}$/$\rm{s}$ to shape the straight and curving portion of the waveguides uniformly. The layer-structured waveguides are distributed in the chip at depths ranging from 70 $\rm{\mu m}$ to 370 $\rm{\mu m}$.

\subsection*{Experiment settings}
The measurement of the fabricated samples is conducted on an optical platform. A continuous-wave (CW) laser with a wavelength of 810 nm is first polarized using a half-wave plate and a quarter-wave plate. The horizontally polarized component is then selected by a polarization beam splitter and coupled into the samples using a 30× input objective lens. The measurement record and intensity distribution of the final state are recorded using a 10× output objective lens and a CCD. For the samples discussed in Sec. \ref{section:manipulation}, an additional lens is employed to capture the full spatial patterns of the final states within the effective region of the CCD.

\section*{Author Contributions}
X. C. performed the experiments and simulations. X. C., R.-Y. Y. and E. B. designed the model of periodic conditional measurement and performed theoretical analysis. X.-W. S. and X. C. designed the waveguide structure to realize conditional measurements. X. C., X.-J. Y., Z.-F. L. and H.-J. D. analyzed the experimental data. X. C. wrote the first version of the manuscript. X. C., R.-Y. Y., X.-W. S., Y.-J. C. and E. B. reviewed and revised the manuscript. X. C., W.-H. Z. and X. T. designed the figures in the paper.
This research is supported by the National Key R\&D Program of China (Grant No. 2024YFA1409300); National Natural Science Foundation of China (NSFC) (Grants No. 62235012, No. 12304342, No. 12574549, No. 12574542, and No. 125B1033); Quantum Science and Technology--National Science and Technology Major Project (Grants No. 2021ZD0301500 and No. 2021ZD0300700); Science and Technology Commission of Shanghai Municipality (STCSM) (Grants No. 2019SHZDZX01, No. 24ZR1438700, No. 24ZR1430700, and No. 24LZ1401500); Shanghai Municipal Commission of Economy and Informatization (Grant No. 2025-GZL-RGZN-02047); Startup Fund for Young Faculty at SJTU (SFYF at SJTU) (Grants No. 24X010502876 and No. 24X010500170); and Frontier Technologies R\&D Program of Jiangsu (Grant No. SBF20250000094). X.-M. J. acknowledges additional support from a Shanghai talent program and support from Zhiyuan Innovative Research Center of Shanghai Jiao Tong University. Hao Tang acknowledges an open research fund from State Key Laboratory of Photonics and Communications and additional support from Yangyang Development Fund.

\bibliography{references}

\end{document}